# Probabilistic reconstruction of global sea surface temperature using generative diffusion models

Haijie Li[1,2], Ya Wang[1,2,3*], Kai Yang[4], Gang Huang[1,2,3*], Xiangao Xia[5*], Ziming Chen[6], Weichen Tao[7], Chenglin Lyu[2], Lin Chen[8,9,10], Miao Zhang[8,9,10], Kaiming Hu[11], Hainan Gong[3], Disong Fu[12], Lin Wang[3]

[1]*Earth System Numerical Simulation Science Center, Institute of Atmospheric Physics, Chinese Academy of Sciences, Beijing, China*

[2]*University of Chinese Academy of Sciences, Beijing, 100049, China*

[3]*State Key Laboratory of Earth System Numerical Modeling and Application, Institute of Atmospheric Physics, Chinese Academy of Sciences, Beijing, China*

[4]*Commonwealth Scientific and Industrial Research Organisation (CSIRO) Environment, Tasmania, Australia*

[5]*Laboratory of Middle Atmosphere and Global Environment Observation (LAGEO), Institute of Atmospheric Physics, Chinese Academy of Sciences, Beijing, China*

[6]*Atmospheric, Climate, & Earth Sciences (ACES) Division, Pacific Northwest National Laboratory, Richland, WA, USA*

[7]*Laboratory of Atmospheric and Oceanic Dynamics, Institute of Atmospheric Physics, Chinese Academy of Sciences, Beijing, China*

[8]*National Satellite Meteorological Center (National Centre for Space Weather), China Meteorological Administration, Beijing 100081 China;*

[9]*Innovation Center for FengYun Meteorological Satellite (FYSIC), China Meteorological Administration, Beijing 100081 China;*

[10]*Key Laboratory of Radiometric Calibration and Validation for Environmental Satellite, China Meteorological Administration, Beijing 100081 China;*

[11]*College of Resource Environment and Tourism, Capital Normal University, Beijing, 100048, China*

[12]*State Key Laboratory of Atmospheric Environment and Extreme Meteorology, Institute of Atmospheric Physics, Chinese Academy of Sciences, Beijing 100029, China*

* Corresponding author: wangya@mail.iap.ac.cn; hg@mail.iap.ac.cn; xxa@mail.iap.ac.cn

**Abstract**

Accurate reconstruction of global Sea surface temperature (SST), which dominates the air–sea coupling and global climate variability, underpins climate monitoring and prediction. Existing SST reconstruction products primarily provide one deterministic field derived from heterogeneous satellite data and in situ observations, limiting their ability to represent observation uncertainty and to support probabilistic forecasting. Here, we introduce Satellite and in situ Adaptive Guided Estimation (SAGE), a diffusion-based uncertainty-aware generative framework for probabilistic SST reconstruction. SAGE learns a physically consistent prior from historical SST data and performs observation-conditioned posterior sampling without requiring satellite or in situ data during training, enabling flexible state inference from heterogeneous observations. Through a progressive data-fusion strategy, observations from two FengYun-3D polar-orbiting satellites constrain basin-scale structures, while sparse in situ measurements serve to refine local anomalies and extremes. The resulting ensemble SST fields well capture observational uncertainty and scale-dependent variability. Validation against independent in situ observations shows that SAGE substantially reduces reconstruction errors compared with widely used operational products. When used to initialize forecasting systems, SAGE-generated SST fields substantially reduce 10-day SST forecast errors relative to current operational analyses. At the climate scale, SAGE-driven forecasts of the 2023–2024 El Niño event show added value in capturing its onset and intensity evolution compared to conventional approaches. Our results demonstrate that SAGE represents a step toward a new paradigm for ocean state estimation and climate prediction.

## Introduction

Sea surface temperature (SST) is a primary boundary condition for the climate system, exhibiting strong variability and dominating ocean–atmosphere heat exchange and global climate (Deser et al., 2010; Frölicher et al., 2018; Mohan et al., 2022). SST variability, such as El Niño–Southern Oscillation (ENSO), can influence hydroclimatic extremes across the globe, including flooding, droughts, and heatwaves in remote regions (Cai et al., 2015; Hobday et al., 2016; McPhaden et al., 2006). Under ongoing greenhouse warming, global mean SST has exhibited a persistent upward trend (Frölicher et al., 2018; Oliver et al., 2018). In 2023, one of the warmest years on record for the global ocean, marine heatwave activity exceeded the historical mean since 1982 by more than three standard deviations, causing severe ecological disruptions and widespread biodiversity loss (Dong et al., 2025a; Dong et al., 2025b). Accurate SST monitoring and data reconstruction are crucial and urgent.

SST products primarily rely on numerical reconstruction, constrained by the heterogeneous nature of the observing system (Kennedy, 2014; Kent et al., 2017). Satellite observations provide near-global coverage but suffer from intrinsic limitations (Donlon et al., 2012; Gentemann et al., 2010; Reynolds et al., 2007). In situ platforms such as drifting and moored buoys deliver highly accurate point measurements but are too sparsely distributed to resolve mesoscale and basin-scale variability (Lindstrom et al., 2012). The coexistence of disparate error characteristics, sampling densities, and spatial scales implies that multiple SST fields can be equally consistent with the available observations. As a result, SST products derived from these heterogeneous measurements are inherently uncertain, yet most existing analyses provide only a single deterministic estimate that does not explicitly represent this uncertainty (Good et al., 2020; Hersbach et al., 2020; Reynolds et al., 2007).

The absence of uncertainty-aware SST estimates limits their utility for probabilistic ocean monitoring and ensemble-based prediction, where quantifying the range of plausible ocean states is essential for forecast reliability and risk assessment (Bauer et al., 2015; Leutbecher and Palmer, 2008). Ensemble representations of initial conditions are known to enhance predictability across time scales, from short-range ocean forecasts to seasonal climate prediction (Buizza, 2018; Palmer et al., 2005). Data assimilation provides a rigorous framework for combining observations with numerical ocean models to generate dynamically consistent analyses (Bouttier and Courtier, 2002; Carrassi et al., 2018; Law et al., 2015), but its application remains computationally demanding and sensitive to model bias, motivating complementary approaches that can deliver fast, observation-driven, and uncertainty-quantified state estimates.

Recent advances in generative deep learning offer a potential alternative for state inference by learning data-driven priors and performing posterior sampling without explicit numerical integration (Ham et al., 2024; Bao et al., 2022; Chao et al., 2025; Chung et al., 2022; Song et al., 2020). Diffusion-based models have demonstrated success in filling cloud-induced gaps in satellite observations and solving noisy inverse problems (Chung et al., 2022; Huber et al., 2024; Malarvizhi and Pan, 2024; Hess et al., 2025), and have recently been coupled with observations to provide data-assimilation-like inference for weather prediction systems (L Huang et al., 2024). Nevertheless, for global SST monitoring, it remains unclear how to construct physically consistent, uncertainty-quantified ensemble SST estimates that faithfully represent heterogeneous observational constraints, and whether such estimates can systematically translate into improved predictive skill across forecasting contexts.

Here, we address these challenges by introducing Satellite and in situ Adaptive Guided Estimation (SAGE), an uncertainty-aware SST reconstruction framework that combines diffusion posterior sampling and a progressive multi-stage fusion strategy for satellite and in situ observations. SAGE explicitly models observation uncertainty and scale-dependent variability while preserving physically consistent SST structures through learned physical priors. By transforming SST reconstruction from deterministic analysis to probabilistic, ensemble-based inference, SAGE provides uncertainty-quantified initial conditions that directly enhance ocean and climate prediction, from short-range SST forecasts to climate-scale phenomena such as ENSO.

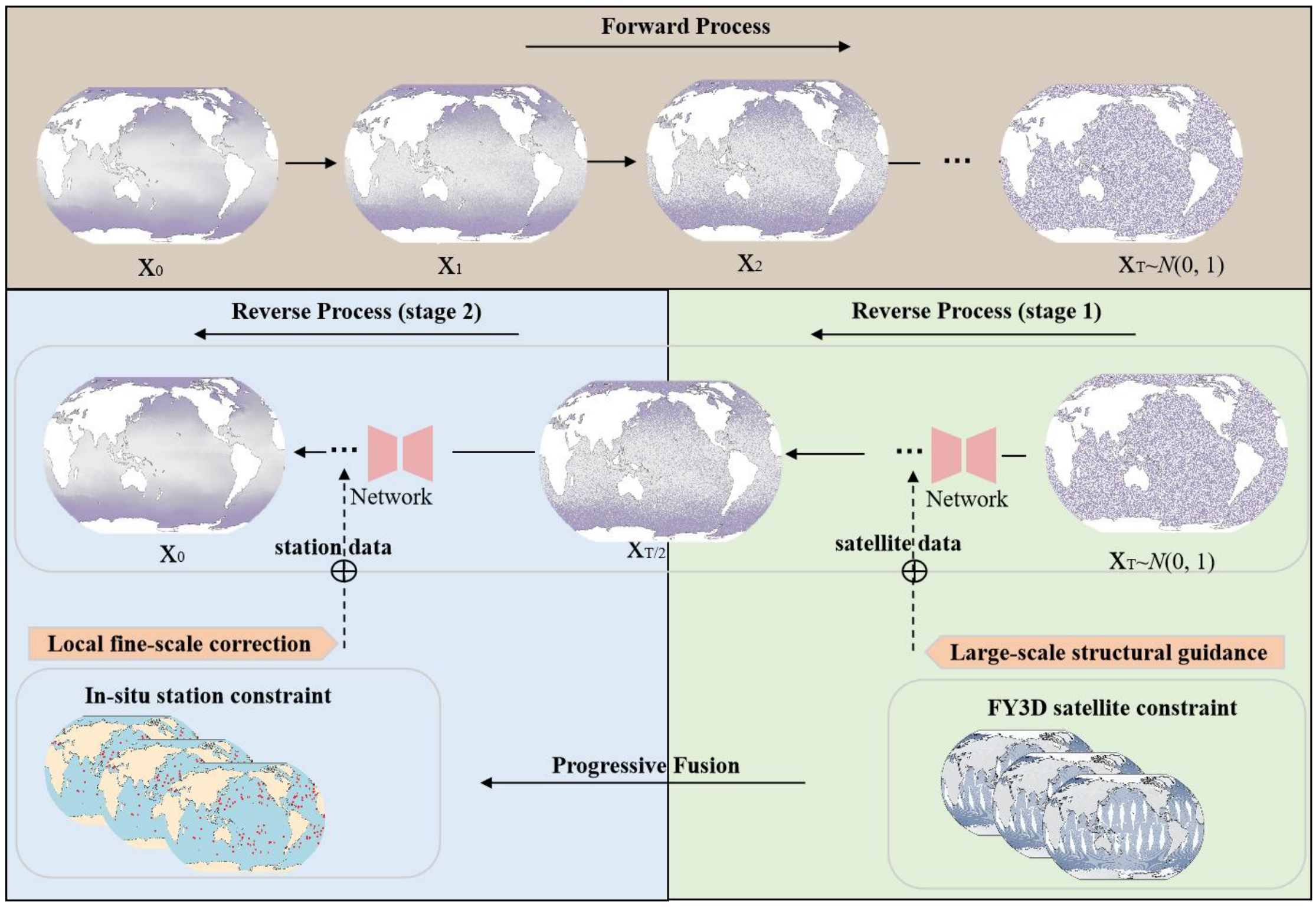

**Figure 1.** Schematic of the diffusion-based SAGE framework. The forward process progressively adds Gaussian noise to an SST field until it becomes a Gaussian distribution. The reverse process samples from this distribution and iteratively removes noise to recover an SST field. To enable multi-source data fusion, the reverse process is divided into two stages: the first stage integrates satellite observations to guide large-scale features, and the second stage incorporates in situ measurements to refine local details.

## Results

### Overview of the SAGE framework

Figure 1 illustrates the diffusion-based SAGE framework for uncertainty-aware

SST reconstruction. SAGE treats SST estimation as a Bayesian inference problem, in which a diffusion model trained on historical SST fields defines a physically realistic prior, and multi-source observations provide constraints on the posterior state. By iteratively sampling from the diffusion reverse process while conditioning on observations, SAGE generates an ensemble of SST realizations that represent the range of ocean states consistent with both the learned prior and the available measurements. Importantly, the resulting ensemble spread reflects posterior uncertainty arising from observational sparsity, noise, and scale mismatch, rather than arbitrary stochastic variability of the generative model.

**Learning a diffusion prior for the historical SST distribution**

Accurate state estimation relies on a physically meaningful prior that captures the system's intrinsic variability (Fowler et al., 2018). We therefore first assess the diffusion model's ability to learn the statistical distribution of historical SST fields from the Optimum Interpolation Sea Surface Temperature (OISST; B Y Huang et al., 2021) reanalysis. After pre-training on the full multidecadal dataset, the model is used to randomly generate 7,300 SST samples, corresponding to approximately 20 years of data. An equal number of SST fields are randomly extracted from OISST to form a reference set. We compare their spatial climatological properties, including the mean, standard deviation, and the 5th and 95th percentiles, to evaluate how well the generated fields reproduce the observed SST distribution.

The diffusion model closely matches OISST in terms of the climatological mean, successfully reproducing the meridional SST gradient with a Root Mean Square Error (RMSE) of 0.16 °C (Figures S1a–b). It also accurately captures regions of elevated variability, such as western boundary currents and coastal zones, achieving an RMSE of 0.13 °C for the standard deviation (Figures S1c–d). Furthermore, the spatial patterns of the 5th and 95th percentiles closely resemble those of OISST, indicating that the diffusion model effectively represents the distribution of extreme SST values (Figures S1e–h). These results demonstrate that the diffusion model provides a statistically consistent prior for SST variability, which is subsequently constrained by observations during posterior inference.

**Global SST reconstruction from satellite and in situ constraints**

Building on the learned prior, SAGE performs global SST reconstruction by iteratively sampling from the diffusion reverse process while conditioning each step on available observations. At each diffusion step, the current sample is mapped to a clean-field estimate, and an observation-consistency term is applied using a gridded observation mask (see Methods, SAGE Model). A tunable guidance strength controls the balance between the learned prior and observational constraints, enabling SAGE to correct systematic biases and fill observational gaps rather than merely interpolating missing values. To fuse heterogeneous data sources, SAGE adopts a progressive strategy in which satellite retrievals constrain early sampling steps to anchor basin-scale structures, followed by stronger in situ constraints at later steps to refine local anomalies and extremes (Figure 1).

To ensure an independent evaluation, reconstruction biases are primarily assessed against the 2023 Met Office Hadley Centre Integrated Ocean Database (HadIOD; Atkinson et al., 2014) in situ database, which is not used during model training. OISST exhibits notable biases along western boundary currents (including the Kuroshio Extension and Gulf Stream), in the mid-to-high latitudes of the Southern Ocean, and in the Arctic, with a global mean RMSE of 0.59 °C (Figure 2a). A satellite-only SAGE reconstruction (SAGE(Sat)) shows broadly similar spatial bias structures, but with a larger overall mismatch (RMSE of 0.72 °C; Figure 2b), suggesting that satellite retrieval errors and sampling gaps can project onto dynamical regimes with strong fronts and eddies. In contrast, the multi-source configuration (SAGE(Sat+Hadi)) substantially reduces biases across these regions and achieves a global mean RMSE of 0.17 °C relative to HadIOD (Figure 2c), indicating that in situ constraints effectively tighten the posterior and suppress regime-dependent satellite biases.

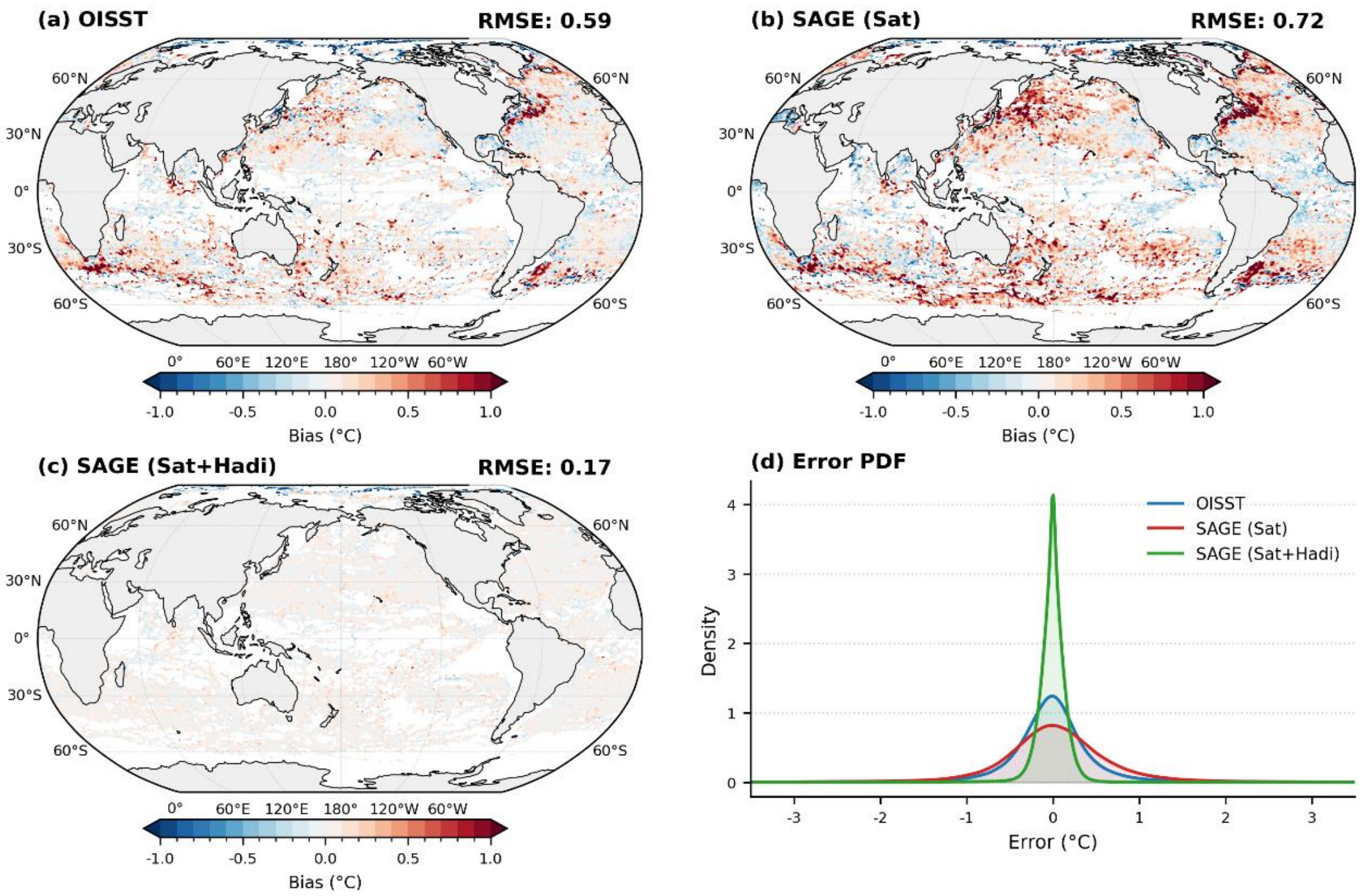

**Figure 2**. Comparison of SST reconstruction biases and error distributions. (a–c) Spatial distributions of SST bias during 2023 for OISST, SAGE(Sat), and SAGE(Sat+Hadi) relative to HadIOD in situ observations, with corresponding RMSE values. (d) Probability density functions (PDFs) of reconstruction errors, where blue, red, and green lines represent OISST, SAGE(Sat), and SAGE(Sat+Hadi), respectively.

Reconstruction skill is further evaluated relative to OISST, which is used here as a widely adopted, spatially complete reference product rather than as ground truth. Raw Fengyun-3D (FY-3D) satellite SST exhibits a global RMSE of 2.10 °C relative to OISST. As a classical baseline, a Navier-Stokes-based inpainting method (Bertalmío et al., 2001) can fill gaps but largely inherits satellite biases, yielding an RMSE of 2.51 °C (Supplementary Figure S2a). In contrast, satellite-constrained SAGE(Sat) substantially reduces the global RMSE to 0.44 °C, though residual errors remain in western boundary current regions (e.g., Kuroshio Extension and Gulf Stream) and parts of the Southern

Ocean (Figure 3a). Adding HadIOD in situ constraints (SAGE(Sat+Hadi)) further reduces the RMSE to 0.30 °C and improves performance across nearly all basins (Figure 3b). An in situ–only ablation yields an RMSE of 0.41 °C (Supplementary Figure S2b), highlighting that sparse point measurements alone cannot provide a spatially coherent optimum and that satellite and in situ observations play complementary roles.

To quantify reconstruction uncertainty, we generate 20-member ensembles for each configuration. Ensemble uncertainty is elevated in dynamically active regions (western boundary currents and eddy-rich areas) and in regions with sparse observational constraints, broadly coinciding with regions of larger deterministic error (Figures 3c and 3d). Consistent with its lower bias, SAGE(Sat+Hadi) exhibits a smaller ensemble spread than SAGE(Sat), indicating that the additional in situ constraints reduce posterior uncertainty. Temporal evolution of reconstructed SST shows that SAGE(Sat+Hadi) closely follows the seasonal cycle represented by OISST throughout the year, while SAGE(Sat) exhibits intermittent deviations during transitional periods (Figure 3e).

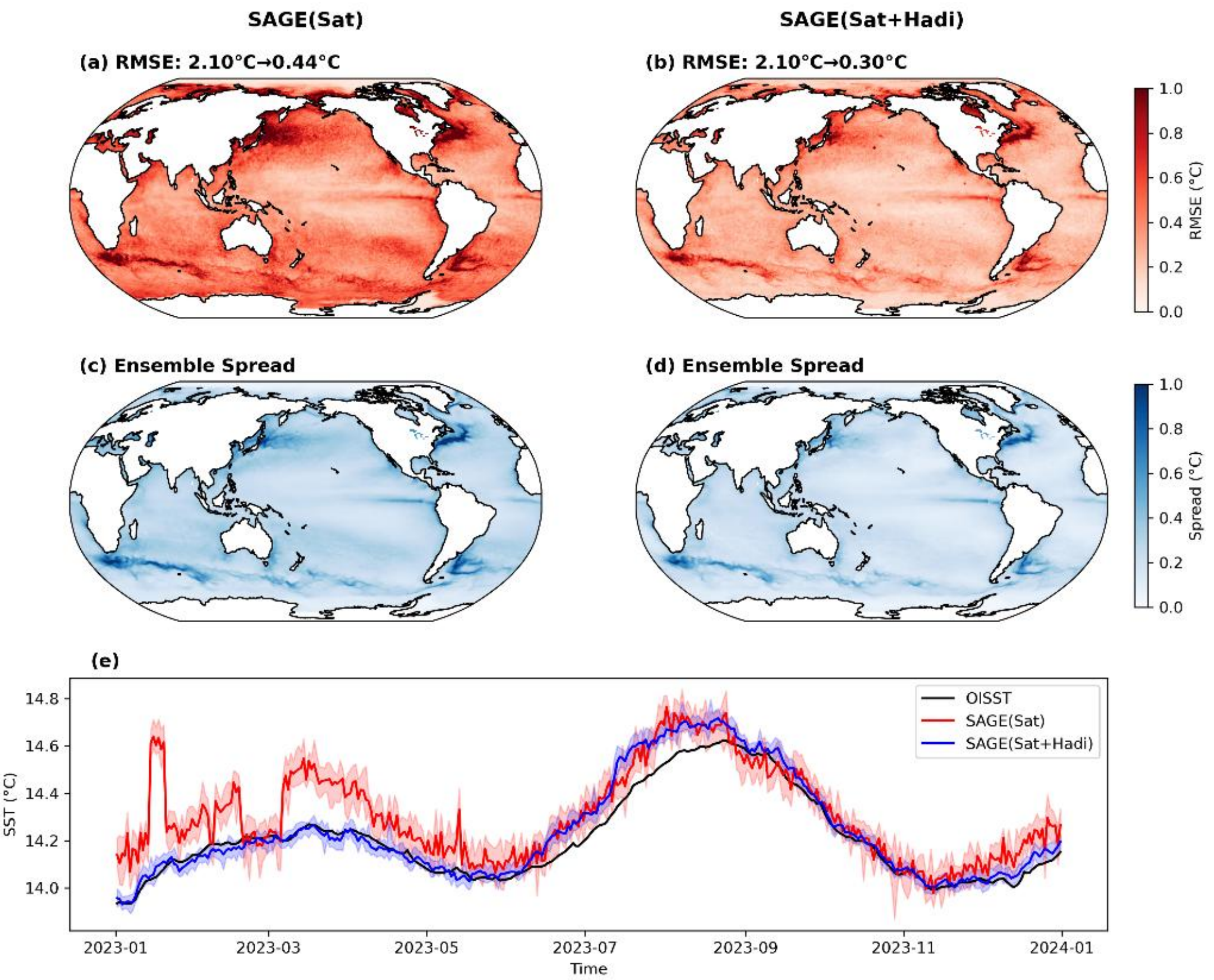

**Figure 3.** Panels (a–b) present the spatial RMSE distributions of SAGE(Sat) and SAGE(Sat+Hadi) relative to OISST in 2023. SAGE(Sat+Hadi) incorporates both satellite and in situ observational constraints, while SAGE(Sat) relies solely on satellite data. Panels (c–d) show the corresponding ensemble uncertainty distributions (20 members), and panel (e) displays the SST time series from OISST, SAGE(Sat), and SAGE(Sat+Hadi).

We evaluate SAGE(Sat+Hadi) against OISST during the record-breaking marine heatwave (MHW) year of 2023 to test its ability to reconstruct SST extremes, focusing on MHW days, maximum duration, and mean intensity. In 2023 summer, both products identify four major MHW centers in the North Pacific, North Atlantic, tropical eastern Pacific, and southwestern Pacific (Figures 4a–c). Outside polar regions, spatial patterns of MHW days are highly consistent, with most differences within ±20 days (Figures 4a–b, S3b). Maximum duration and mean intensity also show close agreement at the basin scale, with typical intensity differences within ~0.1 °C (Figures 4d–h, S3e and S3h). Regional differences are most pronounced during peak-event periods, as revealed by SST anomaly time series within the core of each MHW center (Figures 4j–m). Regions exhibiting larger discrepancies coincide with areas where OISST shows systematic mismatches relative to independent HadIOD observations (Figure 2). Given the absence of a gridded ground truth for daily extremes, this comparison provides a robustness assessment rather than a definitive accuracy evaluation.

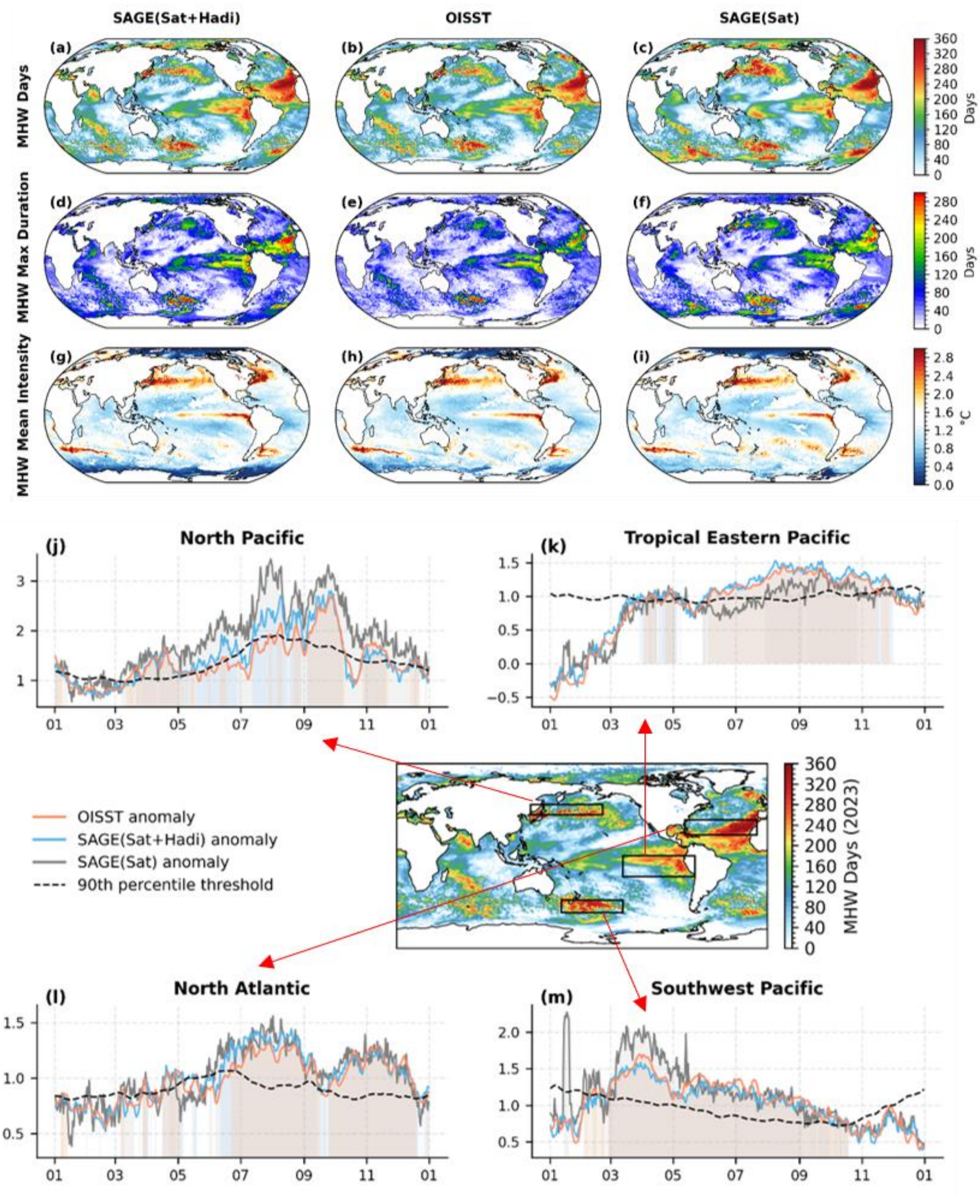

**Figure 4.** Panels (a–c) show the number of MHW days during 2023 for SAGE(Sat+Hadi), OISST, and SAGE(Sat). Panels (d–f) show the longest MHW event durations during 2023 for the same three models. Panels (g–i) show the mean intensity of MHWs during 2023 for the same three models. Panels (j–m) show SST anomaly time series during 2023 for the four selected MHW regions: the North Pacific (40°N~50°N, 120°E~160°W), tropical eastern Pacific (20°S~0°, 140°W~70°W), North Atlantic (20°N~35°N, 80°W~10°W), and southwestern Pacific (55°S~43°S, 160°E~140°W). OISST is shown in orange, SAGE(Sat+Hadi) in blue, and SAGE(Sat) in gray. Dashed lines indicate the thresholds used to define MHW events, and the shaded areas under each solid line indicate periods when MHW events occurred.

## Improved short-range and seasonal prediction using SAGE

Beyond monitoring, reconstructed SST is most valuable when used as an initial/boundary condition for prediction. We therefore assess SAGE in two downstream settings: (1) global SST forecasting at 1–10-day lead times and (2) seasonal climate prediction via SST nudging in a coupled dynamical model (see Methods).

Because the diffusion prior is trained on historical OISST data from 1982–2020, the downstream prediction experiments are evaluated using the independent year 2023 to avoid potential information leakage between training and evaluation. For short-range SST prediction, we train a 10-day SST forecasting model based on a Vision Transformer (Dosovitskiy et al., 2020) architecture using OISST as the training target. To isolate the impact of the initial SST state, all model configurations are kept identical, and only the initialization fields differ. Ensemble forecasts are initialized from SAGE(Sat+Hadi) SST fields and compared against forecasts initialized from OISST, as well as against an operational ocean forecasting system called the Global Ice Ocean Prediction System (GIOPS; Smith et al., 2016). Forecast verification is performed exclusively against independent buoy observations using the Intercomparison and Validation Task Team (IV-TT) framework (Ryan et al., 2015), ensuring that neither training nor initialization data contaminate the evaluation.

Across lead times from 1 to 10 days, SAGE-initialized forecasts consistently outperform both OISST-initialized AI forecasts and GIOPS, exhibiting lower RMSE at all lead times (Figure 5a). Averaged over the 10-day forecast window, SAGE initialization reduces RMSE by 4.90% relative to OISST-based initialization, with improvements persisting throughout the forecast horizon.

To assess impacts at longer time scales, we conduct seasonal prediction experiments in which the SST boundary condition of a coupled climate model is nudged toward SAGE-reconstructed SST, producing 12-month ensemble forecasts initialized on 1 January 2023. These forecasts are compared against an otherwise identical experiment nudged toward OISST. Relative to OISST-nudged forecasts (correlation $r = 0.24$, RMSE = 0.74 °C), SAGE-nudged forecasts exhibit higher skill in capturing the evolution of the 2023 El Niño event, with a stronger correlation ($r = 0.81$) and a lower RMSE (0.30 °C) relative to the observed Niño 3.4 index and reduced amplitude error over the forecast period (Figure 5b). The ensemble mean more accurately reproduces both the timing and magnitude of the warming, while the ensemble spread better reflects forecast uncertainty during the event's growth phase. These results indicate that diffusion-based SST state estimates can deliver tangible gains for operationally relevant seasonal climate prediction, without requiring full ocean data assimilation or additional dynamical constraints.

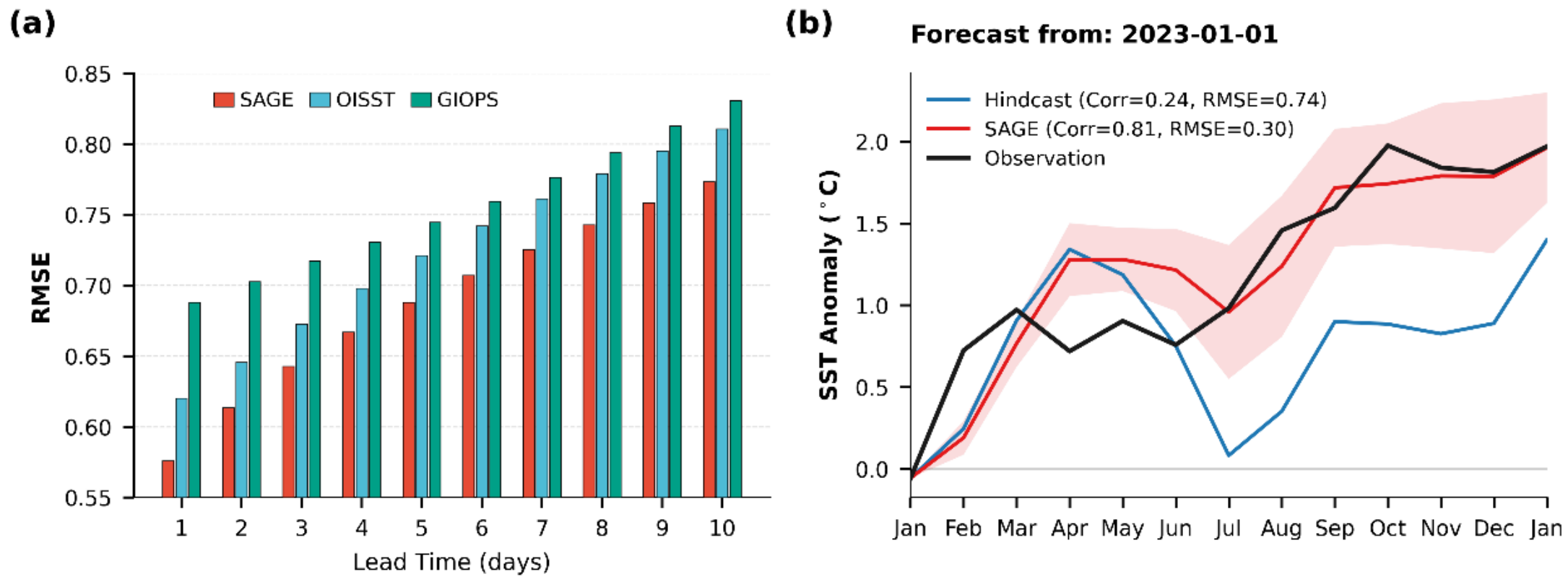

**Figure 5.** Impact of SAGE-reconstructed SST on prediction skill. (a) RMSE of 10-day global SST forecasts verified against independent buoy observations using the IV-TT framework, comparing forecasts initialized from SAGE(Sat+Hadi), OISST, and the operational GIOPS system. (b) Skill of seasonal climate predictions for the Niño3.4 index initialized on 1 January 2023. Red line denotes the ensemble mean of forecasts nudged toward SAGE-reconstructed SST; red shading indicates the full 10-member ensemble spread. Blue line denotes the OISST-nudged forecasts. SST observations are shown in black.

## Conclusion and Discussion

This work demonstrates that global SST reconstruction can be reframed as a probabilistic inference problem in which physically plausible ocean states are sampled from a posterior distribution constrained by heterogeneous observations. By combining a diffusion-based generative prior learned from historical SST variability with explicit observation conditioning, SAGE moves SST estimation beyond deterministic analysis toward uncertainty-aware, ensemble-based state inference, without requiring full numerical data assimilation.

A key feature of SAGE is its ability to integrate disparate observing systems within a unified probabilistic framework. Rather than treating multi-source fusion as a static interpolation problem, SAGE incorporates observations as scale-dependent constraints on the posterior state. In this study, satellite retrievals primarily constrain basin-scale

SST structure, while sparse but high-accuracy in situ measurements progressively refine small-scale gradients and extremes. This principle of progressive constraint integration is general and enables additional observing systems to be incorporated with minimal modification to the underlying framework, while preserving physically consistent SST structures. More practically, the ability of SAGE to retain skill under limited satellite constraints suggests potential for near-real-time SST monitoring, particularly for tracking marine heatwaves and other extremes.

The SAGE-based SST reconstruction ensembles provide practical benefits beyond state reconstruction. When used as initial or boundary conditions, it yields measurable improvements in short-range SST forecasts and seasonal climate predictions, indicating that uncertainty-aware state estimation can enhance predictive skill across time scales. These gains arise without additional dynamical constraints, highlighting the value of probabilistic initial conditions for both data-driven and coupled prediction systems.

Several limitations define the current scope of this work. The present implementation focuses on a single surface variable at a moderate spatial resolution. Extending the framework to multivariate ocean state estimation or finer spatial scales will require larger generative models and increased computational resources, and may benefit from latent-space or hierarchical diffusion formulations. Broader evaluation against independent observing systems and targeted process studies will also be necessary to assess how improvements in SST extremes propagate to air–sea fluxes, atmospheric responses, and coupled predictability. The climate prediction experiments presented here focus on the 2023 ENSO event. While the results highlight the potential benefits of probabilistic SST initialization, future work is required to assess forecast improvements over longer hindcast periods and across multiple climate regimes.

Overall, SAGE provides a scalable pathway for uncertainty-aware SST reconstruction that directly links multi-source observation fusion with probabilistic prediction. More broadly, the framework illustrates how generative posterior sampling can be used to address geoscientific inverse problems characterized by heterogeneous observations and scale-dependent uncertainties, offering a foundation for next-generation, data-driven Earth system state estimation.

## Data and Methods

### Data

Prior training data are taken from the NOAA 1/4° Daily OISST product, which merges satellite, ship, buoy and Argo observations into a gridded climate data record (B Y Huang et al., 2021). All fields are regridded to 0.75° × 0.9375° (240 × 384 pixels). OISST from 1982–2020 is used to train the diffusion prior, while the year 2023 is used for state-inference evaluation.

In situ measurements are from the Hadley Centre Integrated Ocean Database (HadIOD; Atkinson et al., 2014). We retain temperature observations within 0.2 m of the surface and map them onto the 0.75° × 0.9375° grid. For each grid cell, observations within 0.5° are combined using inverse-distance weighting when multiple stations are available.

Satellite observations are obtained from the Microwave Radiation Imager (MWRI) onboard the FY-3D satellite (Chen et al., 2025; Zhang et al., 2018). The dataset used in this study corresponds to the MWRI Level-2 SST orbital product, distributed by the National Satellite Meteorological Center (NSMC), China Meteorological Administration. Both ascending and descending orbital retrievals are used, and their daily averages are computed to estimate daily SST fields. The FY-3D MWRI SST retrievals for 2023 are bilinearly interpolated onto the same grid. Before training and inference, all input variables are linearly scaled to the range [−1, 1] to ensure numerical stability during model optimization.

For the evaluation of short-range SST forecasts, independent observational datasets are used following the Intercomparison and Validation Task Team (IV-TT) framework. To ensure spatial consistency with the model outputs, all evaluation datasets are interpolated onto the same grid prior to verification.

## Methods

### Denoising Diffusion Probabilistic Model

To construct a probabilistic prior for SST fields, we pre-train a diffusion model that learns the data distribution of OISST. The model learns the statistical structure of global SST variability and enables sampling of physically consistent SST fields from latent noise distribution. We adopt the Denoising Diffusion Probabilistic Model (DDPM) framework (Ho et al., 2020; Sohl-Dickstein et al., 2015). During training, the model gradually perturbs SST fields with stochastic noise and learns to recover the underlying signal. The forward process gradually corrupts an OISST field $x_0 \sim p_{\mathrm{OISST}}(x)$ by successively adding Gaussian noise according to a predefined noise schedule $\{\beta_t\}_{t=1}^{T}$:

$$q(x_t \mid x_{t-1}) = \mathcal{N}\left(x_t; \sqrt{1-\beta_t} \cdot x_{t-1}, \beta_t I\right), \quad t = 1, \dots T. \tag{1}$$

After $T$ steps, the sample $x_T$ becomes nearly standard Gaussian. The process has a closed-form solution for any step $t$:

$$q(x_t \mid x_0) = \mathcal{N}\left(x_t; \sqrt{\hat{\alpha}_t} x_0, \sqrt{1-\hat{\alpha}_t I}\right), \tag{2}$$

where $\alpha_t = 1 - \beta_t$ and $\hat{\alpha}_t = \prod_{i=1}^{t} \alpha_i$. Thus, training pairs $(x_t, x_0)$ can be generated without running the full Markov chain.

The generative model aims to invert the noising process and recover an SST field from pure Gaussian noise. The exact reverse transition

$$q(x_{t-1} \mid x_t, x_0) = \mathcal{N}\left(x_{t-1}; \tilde{\mu}_t(x_t, x_0), \tilde{\beta}_t I\right) \tag{3}$$

has mean

$$\tilde{\mu}_t(x_t, x_0) = \frac{\sqrt{\hat{\alpha}_{t-1}}\beta_t}{1-\hat{\alpha}_t} x_0 + \frac{\sqrt{\hat{\alpha}_t}(1-\hat{\alpha}_{t-1})}{1-\hat{\alpha}_t} x_t = \frac{1}{\sqrt{\alpha_t}}\left(x_t - \frac{\beta_t}{\sqrt{1-\hat{\alpha}_t}}\right)\epsilon \tag{4}$$

Where $\epsilon \sim \mathcal{N}(0, I)$ and

$$\tilde{\beta}_t = \frac{1 - \hat{\alpha}_{t-1}}{1 - \hat{\alpha}_t} \quad (5)$$

Because $\tilde{\mu}_t$ depends on the unknown clean field $x_0$, we approximate the process using a neural network that predicts the added noise. Prior work has shown that predicting $\epsilon$ yields more stable training and better generative performance than directly predicting $\tilde{\mu}$ (Ho et al., 2020). We therefore train a noise - prediction model $\epsilon_\theta(x_t, t)$ and compute $\tilde{\mu}_t(x_t, x_0)$ of the reverse Gaussian using equation (4). The variance in the reverse process does not need to remain fixed and can instead be parameterized and learned by the model (Nichol and Dhariwal, 2021). Training minimizes the expected squared error between the true noise and the predicted noise:

$$\mathcal{L}_{\text{simple}} = \mathbb{E}_{t,x_0,\epsilon}\left[\left\|\epsilon - \epsilon_\theta\left(\sqrt{\hat{\alpha}_t}x_0 + \epsilon\sqrt{1 - \hat{\alpha}_t}, t\right)\right\|_2^2\right]. \quad (6)$$

Thus, the model learns how noise was added at each diffusion step, enabling it to remove the noise during sampling.

Once trained, the diffusion model defines a generative prior over SST fields. Starting from random Gaussian noise $x_T \sim \mathcal{N}(0, I)$, the reverse diffusion process sequentially removes noise:

$$x_{t-1} = \mu_\theta(x_t, t) + \tilde{\beta}_t z, z \sim \mathcal{N}(0, I), \quad (7)$$

eventually producing a physically plausible SST field $x_0$. The detailed architectural specifications of $\epsilon_\theta(x_t, t)$ can be found in Figure S4.

The model was trained using OISST data from 1982 to 2020, with a batch size of 24 and the AdamW optimizer at a learning rate of $1 \times 10^{-4}$. The noise schedule was chosen to be linear, with $\beta_t$ increasing from 0.0001 to 0.02 over the course of T steps. The training process ran for 400,000 steps across four NVIDIA GeForce RTX 4090 GPUs, after which the final trained model was saved for subsequent sampling and inference.

## SAGE Model

For state reconstruction, observational constraints are incorporated directly during the generative sampling process. In this framework, SST estimation is formulated as an inverse problem in which the generative model provides a prior distribution of physically plausible SST states, while observations act as partial constraints on the posterior solution. Such formulations have recently been explored in generative modeling approaches, where learned priors are combined with observational information to guide the reconstruction of high-dimensional geophysical fields (e.g., Chung et al., 2022; Bertalmio et al., 2000; Ongie et al., 2020; Song et al., 2020). The general form of this problem can be expressed as:

$$y = A(\boldsymbol{x}_0) + n, y, n \in \mathbb{R}^n, x_0 \in \mathbb{R}^d , \tag{8}$$

where $A(\cdot)$ is a linear degradation operator, $x_0$ represents the original image, y is the degraded image, and $n$ is white noise.

Within the diffusion framework, the reconstruction is performed through a conditional sampling procedure that iteratively refines SST realizations while enforcing consistency with the available observations. In practice, the observation operator $A(\cdot)$ is implemented using a mask derived from the gridded observational coverage, where observed locations are assigned unit weights and unobserved locations are ignored. After each sampling step in the diffusion process, we compute the inferred clean image $\hat{x}_0$ via equation (2) from $x_t$. Then, the constraint term is computed as follows:

$$\nabla_{x_t} \| y - A(\hat{x}_0) \|_2^2 . \tag{9}$$

This term measures the degree of similarity between $A(\hat{x}_0)$ and the observed data $y$. The closer the two are, the smaller the constraint term becomes, indicating less correction is needed. Conversely, the further apart they are, the larger the constraint term, signaling that more correction is required. The algorithm for this constrained sampling process is as follows:

---

**Algorithm:** SAGE

---

**Require:** $T, y_{\text{Sat}}, y_{\text{Hadi}}, \{\zeta_t\}_{t=1}^{T}, \{\tilde{\beta}_t\}_{t=1}^{T}$,

1: $x_T \sim \mathcal{N}(0, I)$

2: **for** $t = T - 1$ **to** 0 **do**

3: **if** $t < \frac{T}{2}$ **then**:

4: $y \leftarrow y_{\text{Hadi}}$

5: **else**:

6: $y \leftarrow y_{\text{Sat}}$

7: **endif**

8: $\epsilon \leftarrow \epsilon_\theta(x_t, t)$

9: $z \sim \mathcal{N}(0, I)$

10: $x'_{t-1} \leftarrow \frac{1}{\sqrt{\alpha_t}}\left(x_t - \frac{\beta_t}{\sqrt{1-\hat{\alpha}_t}}\right)\epsilon + \tilde{\beta}_t z$

11: $\hat{x}_0 \leftarrow \frac{1}{\sqrt{\hat{\alpha}_t}}(x_t + (1 - \hat{\alpha}_t)\epsilon)$

12: $x_{t-1} \leftarrow x'_{t-1} - \zeta_t \nabla_{x_t} \|y - A(\hat{x}_0)\|_2^2$

13: **end for**

14: **return** $x_0$

During the entire inference process, the batch size was set to 8, and the number of steps, T, remained fixed at 1000. For the SAGE(Sat+Hadi) model, the parameter $\zeta_t$ was set to 0.3 for the first 500 steps and to 3.0 for the remaining 500 steps. For the SAGE(Sat) model, the entire process used $y \leftarrow y_{\text{Sat}}$, with $\zeta_t$ set to 0.3 throughout. For the SAGE(Hadi) model, the entire process used $y \leftarrow y_{\text{Hadi}}$ with $\zeta_t$ set to 6.0. These parameter values were selected based on their optimal performance during the inference phase.

**Short-range SST forecasting experiment**

We train a data-driven SST forecasting model using a Vision Transformer architecture (Han et al., 2023) to predict daily SST fields up to 10 days ahead. The model consists of eight identical, sequential encoders. Each encoder features alternating layers of Multi-head Self-Attention (MSA) and Multi-Layer Perceptron (MLP) blocks, with Layer Normalization applied before each block and residual connections after. The MLP comprises two Fully Connected Layers utilizing GELU non-linear activations. Prior to encoding, the initial SST fields are partitioned into 16 x 24 patches, followed by linear projection and positional encoding. The forecast for the subsequent day is then reconstructed from the final encoder's output via linear projection and patch recovery.

Training was conducted on historical OISST (1982–2020) for 600 epochs (batch size: 64; learning rate: 0.0002) and evaluated in 2023. Forecasts were initialized using OISST and SAGE (Sat+Hadi), with ensemble runs generated by sampling multiple initial SST realizations from the SAGE posterior. Model performance was verified against independent buoy observations following the IV-TT evaluation framework.

**Coupled climate prediction with SST nudging**

We conducted 12-month ensemble climate predictions using a fully coupled general circulation model, ICM (Liu et al., 2018). The model integrates the ECHAM6.3 atmospheric component (Stevens et al., 2013) with the NEMO3.6 ocean engine (Madec, 2015).

Model initialization was performed by assimilating the NOAA 1/4° Daily OISST dataset for the period from January 1, 1982, to December 31, 2022. A simple nudging scheme was used, whereby applied heat fluxes nudged the model SST toward the observed values (Auroux and Blum, 2008; Keenlyside et al., 2008; Luo et al., 2005). The temperature feedback magnitude was set to -600 W $m^{-2}$ $K^{-1}$ throughout, corresponding to a relaxation timescale of approximately 2 days, ensuring that the model state is closely constrained to the observed trajectory.

Following initialization, we performed a set of control forecasts. To investigate the sensitivity of forecast skill to initial condition uncertainty, two experimental designs

were employed:

(1) Control Forecast: The nudging toward NOAA OISST was extended through January 31, 2023, to generate a single, deterministic initial state. From this state, a 12-month free forecast was integrated from February 1, 2023, to January 31, 2024.

(2) Ensemble Forecast with SAGE Initialization: To account for initial condition uncertainty, we replaced the NOAA OISST with ten distinct SST realizations from the SAGE (Sat+Hadi) posterior. Consistent with the Control Forecast run, these ten realizations were independently assimilated from January 1 to January 31, 2023, via the same nudging procedure. This yielded a 10-member ensemble of initial states, from which a 12-month ensemble forecast was initiated for the identical target period.

Finally, the prediction skill for the 2023 El Niño event was evaluated using standard indices (e.g., the Niño-3.4 index) and spatial pattern metrics.

**Data Availability**

Daily sea surface temperature data from the OISST product are available from NOAA at https://www.ncei.noaa.gov/products/optimum-interpolation-sst. In situ ocean observations from the HadIOD can be accessed at the Met Office website: https://www.metoffice.gov.uk/hadobs/hadiod/download-hadiod1-2-0-0.html. Buoy observations and GIOPS forecast system data from the IV-TT framework are available at https://thredds.nci.org.au/thredds/catalog/rr6/intercomparison_files/catalog.html. FY-3D MWRI SST Level-2 data are available from the National Satellite Meteorological Center (NSMC), China Meteorological Administration: https://satellite.nsmc.org.cn/DataPortal/cn/data/detail.html.

**Code Availability**

The codes are available from corresponding author upon reasonable request.

**Acknowledgements**

This work is supported by the National Natural Science Foundation of China (Grant Nos. 42141019, 42175049, and 42261144687) and the Second Tibetan Plateau

Scientific Expedition and Research (STEP) program (Grant No. 2019QZKK0102). We also acknowledge the support and resources provided by the National Major Scientific and Technological Infrastructure " Earth System Science Numerical Simulator Facility" (EarthLab).

## Author Contributions

Y.W. conceived the study and designed the experiments. H.L. performed the model simulations, conducted the data analysis, and wrote the initial draft of the manuscript. Y.W. and G.H. supervised the project and provided primary intellectual guidance. C.L. performed the nudging experiments. X.X. and K.Y. contributed to the results interpretation and the revision of the manuscript. Z.C. and W.T. contributed to the discussion of the results. L.C., M.Z., K.H., H.G., D.F., and L.W. provided technical support and assisted with the manuscript revision. All authors discussed the results and commented on the final version of the manuscript.

## Competing Interests

The authors declare no competing interests.